\documentclass[aps, prd, twocolumn, showpacs]{revtex4-1}
\usepackage{dcolumn,amsmath,amssymb,graphicx}

\begin{document}

\title{Comment on ``Astrophysical gravitational waves in conformal gravity"}

\author{F. F. Faria}
\email{felfrafar@hotmail.com}
\affiliation{Centro de Ci\^encias da Natureza, Universidade Estadual do 
Piau\'i, 64.002-150 Teresina, PI, Brazil}


\begin{abstract}  
Recently, in Phys. Rev. D \textbf{98}, 084002 (2018), an investigation was 
carried out on the gravitational radiation from binary systems in conformal 
gravity and massive conformal gravity. According to the authors 
of the paper, these two theories differ only by the sign of a parameter in 
the matter action of a general conformal theory. For a negative 
parameter the general conformal theory indeed corresponds to conformal gravity. 
However, here, we argue that for a positive parameter the general conformal theory 
does not correspond to massive conformal gravity.  
\end{abstract}

\maketitle

As addressed in Ref. [70] of Ref. \cite{Caprini}, the massive conformal 
gravity has one scalar field (dilaton field) coupled to the 
gravitational part of the theory \cite{Faria1} and a second scalar field 
(Higgs field) coupled to the fermionic matter part of the theory 
\cite{Faria2}. It is then stated in Ref. [70] of Ref. \cite{Caprini} that, after 
the imposition of a Weyl gauge, the massive conformal gravity corresponds 
to the general conformal theory with a positive parameter considered in Ref. 
\cite{Caprini}. This is indeed true if the Weyl gauge is imposed on the 
massive conformal gravity action. However, this procedure is not suit for 
the gauge analysis of the theory. By imposing the Weyl gauge on the massive 
conformal gravity field equations, we find $R=0$. Since this condition does 
not hold in the general conformal theory with a positive parameter \cite{Foot}, 
then the two theories do not correspond to each other.

\section*{Acknowledgments}
The author would like to thank the authors of Ref. \cite{Caprini}, in 
particular P. H\"olscher, for useful discussions.



\begin{thebibliography}{99}
\bibitem{Foot}
The authors of Ref. \cite{Caprini} have suggested calling this theory 
``conformal Einstein-Weyl-Higgs gravity" for future reference, with which I 
agree.
\end{thebibliography}
\end{document}